\documentclass{article}
\usepackage[english]{babel}
\usepackage{authblk}
\usepackage[letterpaper,top=2cm,bottom=2cm,left=3cm,right=3cm,marginparwidth=1.75cm]{geometry}
\usepackage{amsmath}
\usepackage{graphicx}
\usepackage[colorlinks=true, allcolors=blue]{hyperref}
\usepackage{caption}
\usepackage{subcaption}
\usepackage{csquotes}
\usepackage[backend=biber,style=numeric,maxnames=10, minnames=1,  maxcitenames=10]{biblatex}
\usepackage{orcidlink}       
\title{Construction techniques and commissioning of the Three-Backlink Experiment for the LISA mission}
\author{Lea Bischof$^{1,2,*}$\orcidlink{0000-0003-1576-7274}, Melanie Ast$^{1,2}$\orcidlink{0000-0003-1911-1686}, Jiang Ji Ho-Zhang$^{1,2}$\orcidlink{0000-0003-0379-8997},
        Nicole Knust$^{1,2}$\orcidlink{0000-0002-5984-5353}, Daniel Penkert$^{1,2}$\orcidlink{0000-0001-6120-7118}, Daniel Jestrabek$^{1,2}$\orcidlink{0009-0004-9795-9388}, Jens Reiche$^{1,2}$\orcidlink{0009-0001-1147-9366},
        Thomas S. Schwarze$^{3}$\orcidlink{0000-0002-2260-9971}, Katharina-Sophie Isleif$^{4}$\orcidlink{0000-0001-7032-9440}, Oliver Gerberding$^{5}$\orcidlink{0000-0001-7740-2698}, 
         Gerhard Heinzel$^{1,2}$\orcidlink{0000-0003-1661-7868}, Stefan Ast$^{3}$\orcidlink{0000-0003-4531-273X}, Karsten Danzmann$^{1,2}$\orcidlink{0000-0003-4174-683X}}
\affil{$^1$Max-Planck-Institut für Gravitationsphysik (Albert-Einstein-Institut), Hannover, Germany}
\affil{$^2$Leibniz Universität Hannover, Hannover, Germany}
\affil{$^3$Deutsches Zentrum für Luft-und Raumfahrt e.V., Hannover, Germany}
\affil{$^4$Helmut Schmidt Universität, Hamburg, Germany}
\affil{$^5$Universität Hamburg, Hamburg, Germany}
\affil{$^*$Author to whom any correspondence should be addressed.}
\begin{document}
\maketitle
\begin{abstract}
Designed to detect gravitational waves in the lower-frequency band, the space mission LISA will open a new window to astronomy after its launch in the 2030s.
Each LISA spacecraft houses two optical benches that require the exchange of a phase reference between them via an optical connection, called a Backlink.
Here we present the construction and commissioning of an ultra-stable quasi-monolithic optical testbed to investigate different Backlink implementations: a direct fiber, a frequency-separated fiber, and a free-beam link, compared in the Three-Backlink Experiment. Dedicated alignment techniques crucial for the construction of these optical benches are presented together with the development of a high-precision beam alignment and measurement tool - a Calibrated Quadrant Photodiode Singleton.
An upper limit for the performance of all three investigated Backlink schemes, as determined by initial experiments, can be set at a $15\text{pm}/\sqrt{\text{Hz}}$-equivalent level within the LISA band, spanning 0.1mHz to 1Hz.
Our measurements were able to verify the successful construction and commissioning of this very complex interferometer as an interferometric laboratory testbed for LISA. We find no limitations due to the construction on the here reported performance levels. Our results can support the construction of high-precision metrology testbeds for space-based laser interferometry for future gravitational wave or geodesy missions.
\end{abstract}
\section{Introduction}
The Laser Interferometer Space Antenna (LISA) will detect gravitational waves in the frequency band between 0.1\,mHz to 1\,Hz. LISA uses long-range laser interferometers to measure the displacement between free-floating test masses and, by that, the imprinted gravitational wave signal.
The test masses are housed in three spacecraft that are separated by 2.5 million kilometers and form a triangle constellation that performs a cartwheel orbit around the Sun, following the Earth.\\
The primary challenge relevant to this paper is the breathing angle of $\pm1.5$° between the laser links, which is caused by orbital dynamics.
Compensation of this angular motion is studied in two concepts, the in-field pointing \cite{Hasselmann2021}, and the LISA baseline of telescope pointing \cite{LISAredbook}.
In the baseline design of telescope pointing, movable subassemblies within each spacecraft compensate for angular motion, thereby guaranteeing the continued interference of the far spacecraft laser beam with the local laser beam.
Each spacecraft houses two subassemblies and thus two lasers and two optical benches. To connect the laser links between all arms, especially motivated by the TDI algorithm to suppress laser frequency noise \cite{Tinto2004TDI,Shaddock2003TDI}, an optical connection between the two optical benches in each spacecraft is required. This connection is called the LISA Backlink \cite{LISAredbook}. It is often also referred to as the Phase Reference Distribution System (PRDS), which addresses the purpose of exchanging the phase information between the optical benches.\\
The Backlink is a bi-directional link, which provides the critical feature of suppressing any phase disturbances that are equal in both directions, a property known as reciprocity. Non-reciprocities will create excess noise and degrade the mission performance. Hence, the Backlink is required to achieve a non-reciprocal pathlength stability in the order of  $1\frac{\text{pm}}{\sqrt{\text{Hz}}}$.\\
The experiment presented here provides a testbed for investigating this LISA Backlink. While it was developed in the context of the LISA mission, it also has high relevance for other LISA-like space missions. These include several detector designs in earlier development stages, like Taiji \cite{Lou2020Taiji}, TianQin \cite{XuPaper2025TianQinBacklink}, or potential other future formations flying missions with laser interferometers along more than one arm \cite{Kawamura2011Decigo,Crowder2005BeyondLISAmissions,Sesana2021}.\\
Several investigations on the Backlink have been performed in the past, and their key findings are summarized below.
An experiment with a fiber Backlink \cite{FleddermannPhD} revealed backscatter of the Backlink fiber and its coupling into the measurement as a critical noise contribution.
Because backscatter becomes more critical with increased phase dynamics, the backscatter in the input fibers, which are exposed to comparable low temperature stability in the laboratory, also has high relevance. This effect couples less in the actual LISA optical bench due to a large attenuation and higher thermal and vibrational stability compared to this laboratory experiment.\\
Deeper studies of fiber performances were done in \cite{RohrPhD},\cite{Rohr2020}, using single fibers, not implemented in a Backlink configuration. These investigations present an understanding of the temperature-to-phase coupling and the levels of backscatter under radiation exposure.\\
The first experiment mentioned above was built onto one single optical bench, which enabled the subtraction of differential laser phase noise via a reference interferometer. 
Experiments over two benches connected via a Backlink in the form of a pathlength-controlled free beam were tested in \cite{Isleif2018} and \cite{ChiltonPhD}. They demonstrated the feasibility of keeping the interferometric contrast during a rotation between the benches. However, these experiments lacked the aforementioned reference for differential laser phase noise subtraction and could thus not probe the performance of the reciprocity to the required pm-level.\\
The Three-Backlink Experiment (3BL) was designed based on those previous findings. It consists of two separate optical benches, connected by three different Backlink implementations: a direct fiber, a frequency-separated fiber implementation, and a free-beam Backlink.
It provides the option to rotate the benches, and by using more than one Backlink implementation, enables the subtraction of differential laser phase noise \cite{Isleif2018}. By choosing three connections, the comparison of three different measurements enables the disentangling of the noise behavior of each individual implementation.\\
In the following sections, we introduce the different Backlink implementations, their design approaches to minimize their non-reciprocal phase noise, and the method of probing the latter to the desired performance level. We then focus on the construction of the two ultra-stable benches, introduce a tool for high-precision beam position measurements, and discuss our construction strategies.
We approached the challenges of achieving interference contrast while optimizing fiber-to-fiber coupling, controlling beam parameters, and minimizing tilt-to-length effects by precisely overlapping two counter-propagating beams.
We conclude on the construction by characterizing the finalized benches in terms of interference contrast, fiber coupling efficiency, beam parameters, and vertical component tilts.\\
In the last part, we present the resulting noise levels after the initial commissioning phase and our understanding of the limiting noise sources. Based on those, we set an upper limit for the non-reciprocity of each Backlink implementation.
\section{The Three-Backlink Experiment}
\begin{figure}[ht]
    \centering
    \includegraphics[width=0.95\linewidth]{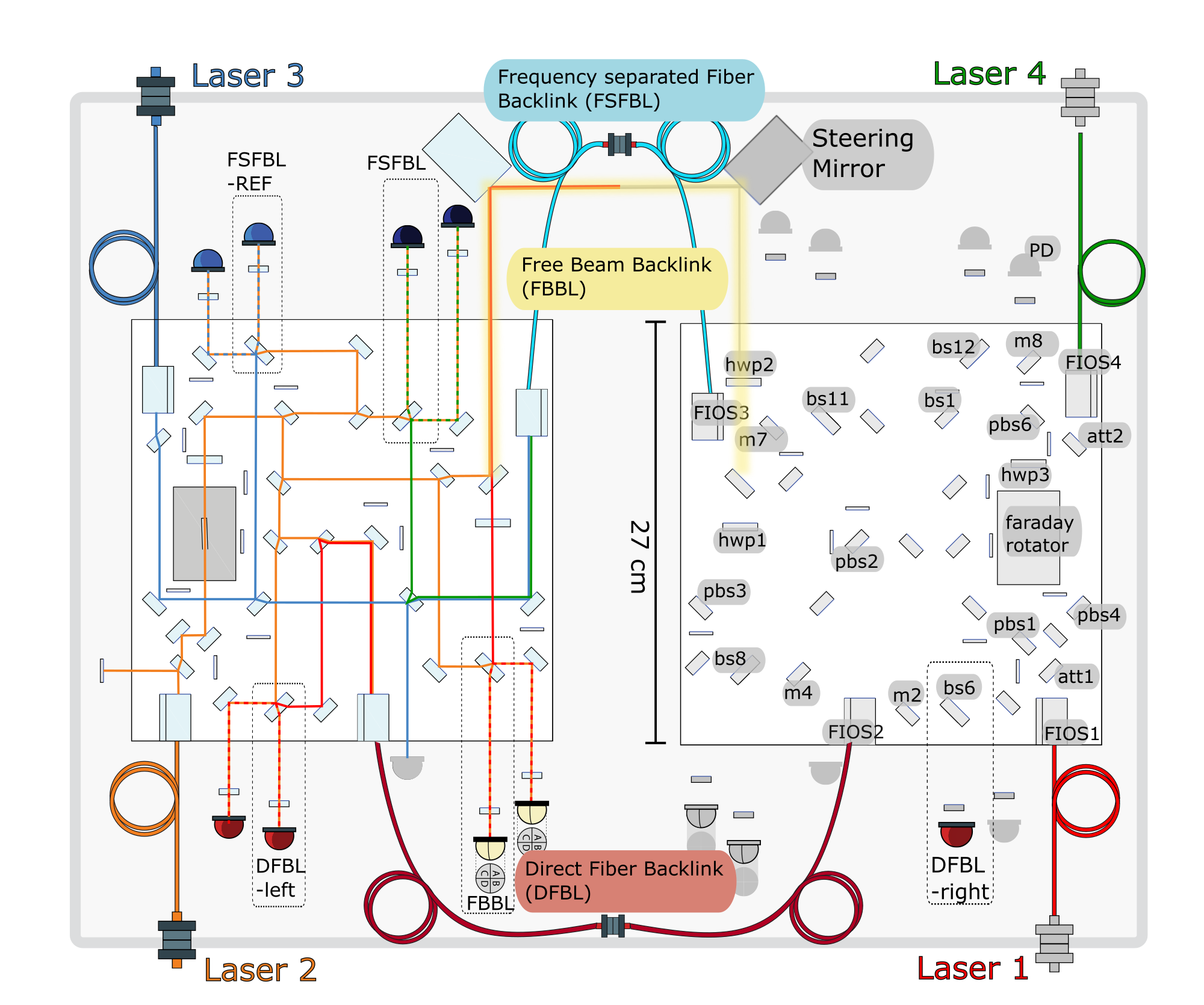}
    \caption{
    The 3BL consists of two optical benches with a mirrored layout of the optical components, except for one additional half-wave plate on the right bench. On the left side, the bench is shown with the focus on the beam paths, while the right side presents the optical design with labeled components:
    bs: beamsplitter, m: mirror, pbs: polarizing beamsplitter, hwp: half-waveplate, att: attenuator, FIOS: fiber injector optical subassembly. Some optics are not labeled for simplicity. This figure is based on \cite{IsleifPhD}, originally created from IfoCAD simulations \cite{Kochkina2013IfoCAD} and with a swap of the left and right bench.}
    \label{fig:3BLoverview}
\end{figure}

Based on the summarized findings above, an experiment was designed that consists of two separate optical benches, has a reference for differential laser noise subtraction, and incorporates three Backlinks with different strategies for mitigating backscatter contributions. These Backlink schemes were chosen from a variety of potential Backlink designs, presented in \cite{IsleifPaper2017} and \cite{IsleifPhD}, and include two fiber-based and one free-beam implementation.
In a sophisticated design process, the layout as presented in figure\,\ref{fig:3BLoverview} was found in \cite{IsleifPhD} by focusing on minimizing effects of spurious beams and enabling measurements with pm-precision.
The design is described in detail in \cite{Isleif2018}, and a short overview will be given in the following.\\
Both benches are mirror-symmetrical, inject two laser beams from the laser preparation, exchange three laser beams, and house four interferometers each. The design combines all three Backlink connections in one experiment, using the same main lasers.
To isolate the noise contributions from the Backlinks during our performance measurements, a Faraday isolator and an attenuator are placed in front of the input fibers to minimize their backscatter influence. While LISA will not have a Faraday isolator on its optical benches, its telescope pick-off has an even higher power splitting ratio than the attenuator used here and thus reduces backscattering from the input fibers significantly.
\subsubsection*{The Direct Fiber Backlink (DFBL)}
The DFBL is highlighted in red in figure \ref{fig:3BLoverview} and exchanges laser light from Laser 1 and Laser 2 between the two benches.
It only consists of an optical fiber connecting the two benches and is thus the most straightforward choice for an optical connection.
Due to the known noise contribution from backscatter, the baseline for LISA foresees a direct fiber connection with attenuation strategies that reduce the relative power sent to the Backlink fiber. 
The DFBL in the 3BL is not using such an attenuation and thus has a higher and thereby easier to study scatter contribution.
\subsubsection*{The Frequency Separated Fiber Backlink (FSFBL)}
In figure\,\ref{fig:3BLoverview}, the FSFBL is highlighted in blue. It is also a fiber connection, but uses two additional lasers to exchange the phase information between the benches.
In the figure, they are labeled Laser 3 and Laser 4, with frequencies $f_3$ and $f_4$.
On the adjacent bench, they interfere with the local beam from Laser 1 or 2, respectively. One additional interferometer per bench, labeled FSFBL-REF, is needed to monitor and subtract the differential phase noise with respect to the local lasers.
With this design, the limitation by backscatter noise from the fiber is circumvented by separating the frequency of the backscattered light from the measurement frequencies.
However, the need for two additional lasers and interferometers significantly increases the design's complexity, making it less appealing for a space mission. The complexity grows even further with the control and read-out electronics for these additional lasers and channels.
\subsubsection*{The Free Beam Backlink (FBBL)}
This Backlink scheme is shown as yellow beam path in figure\,\ref{fig:3BLoverview}.
A free-space beam is guided by two steerable mirrors, controlled by two Differential Wavefront Sensing (DWS)-based control loops, which maintain the beam orientation and thus the reciprocal connection between the benches.
Accordingly, this Backlink scheme eliminated the fiber backscatter completely. However, it introduces additional challenges, like increased phase dynamics of spurious beams caused by the motion of the mirrors. For LISA, it furthermore necessitates the optical and mechanical control system of the steering mirrors and the routing of a free beam between the benches and their movable optical subassemblies.
\subsection{Non-reciprocal phase information}
The key relevant parameter of the here presented experiment is the reciprocity of the Backlink connection, with a requirement of the non-reciprocal noise in the order of  $1\frac{\text{pm}}{\sqrt{\text{Hz}}}$.
This noise is extracted by sending light one-way through the Backlink, interfering it with the local laser beam, and then comparing the interferometric phase to that of the counterpart interferometer on the adjacent bench. The difference between the two phases includes the differential phase shift created by the Backlink.
While the physical delays of the counter-propagating laser beams may be identical in both directions, we are particularly concerned with measuring noise sources, such as scattered light.\\
The non-reciprocal phase noise extraction is presented in the following for the example of the DFBL.
In a perfectly reciprocal case, the light traveling in one direction would experience the exact same phase behavior as light traveling the Backlink connection in the opposite direction. Following the non-reciprocal phase description in \cite{IsleifPhD}, this results in:
$$\varphi_{\rightarrow}-\varphi_{\leftarrow}=0.$$
To extract the non-reciprocal phase contribution in the DFBL, we combine the phases measured at the two PDs labeled in figure\,\ref{fig:3BLoverview} by $\text{DFBL}_\text{left}$ and $\text{DFBL}_\text{right}$. We assume that Laser\,1 has a higher frequency than Laser\,2 and thus positive phase shifts on Laser\,2 appear as negative phases in their combinations:
\begin{equation}
    \label{eq:non-recDFBL}
    \Phi_\text{DFBL} = \text{DFBL}_\text{left}+\text{DFBL}_\text{right} = \varphi_{\rightarrow}-\varphi_{\leftarrow} + 2\cdot(\varphi_{\text{Laser1}} - \varphi_{\text{Laser2}})    
\end{equation}
Additional terms of laser frequency noise that couple via unequal armlength in the interferometers cancel in this non-reciprocity combination because we assume perfect symmetry of the optical benches.\\
The first part of equation\,\ref{eq:non-recDFBL} is the to-be probed non-reciprocal phase noise related to the Backlink. The noise is dominated by the second term ($2\cdot(\varphi_{\text{Laser1}} - \varphi_{\text{Laser2}})$), which is twice the differential noise of the two involved lasers.\\
A combination as in equation\,\ref{eq:non-recDFBL} is performed for all three Backlink implementations.
It is identical for the FBBL and more exhaustive for the FSFBL, where two interferometers per bench are combined with their counterparts. See \cite{IsleifPhD} for more details on the calculation for each Backlink.
All these combinations are limited by the same differential laser noise, which thus cancels in pairwise combinations. Accordingly, we construct the following combinations:
\begin{equation}\label{eq:All3BLsEquation}
    \begin{split}
        &\Phi_\text{DFBL-FBBL}=\Phi_\text{DFBL}-\Phi_\text{FBBL}\\
        &\Phi_\text{FBBL-FSFBL}=\Phi_\text{FBBL}-\Phi_\text{FSFBL}\\
        &\Phi_\text{FSFBL-DFBL}=\Phi_\text{FSFBL}-\Phi_\text{DFBL}
    \end{split}
\end{equation}
These three combinations will be extracted from the measurements discussed in the Results section. They each consist of the combined non-reciprocities of two Backlinks, and through comparison, we can study the performance of each individual Backlink.
In order to perform said measurements to the required pm-level, the optical benches that form the testbed must be constructed with the appropriate stability. We will address this in the following section.
\section{Construction}
Measurements of the Backlinks' phase and the equivalent pathlength stability down to a pm level within the LISA measurement bandwidth require a testbed formed by ultra-stable optical benches.
Our base material is the low-CTE Clearceram (by OHARA GmbH), where the fused-silica components are quasi-monolithically attached using UV-curable glue.
Comparable optical bench construction techniques, such as hydroxide-catalysis bonding or optical contacting, guarantee a better perpendicularity and the former for a much higher durability, but have the disadvantages of short alignment windows and high complexity \cite{Elliffe2005}.
Meanwhile, the UV epoxy Optocast 3553 is sufficiently stable \cite{GerberdingPaper2017}, and allows for alignment times up to 6\,h, after which it is cured with UV-light at a wavelength of 365\,nm.\\
More details of the construction techniques can be found in \cite{BischofPhD} and \cite{KnustMaster}.
\subsection{General construction challenges}
The optical bench layout as presented in figure\,\ref{fig:3BLoverview} is populated with a high number of components that create interferences in four interferometers and two fiber-to-fiber couplings per bench. Both require a good mode overlap and thus fine alignment of each of the two beams involved.\\
The optical benches and the optics' bottom surface are polished to provide good flatness and perpendicularity to the optical surface. However, by working with glue, the intrinsic assumption of a beam sufficiently plane-parallel to the optical bench does not hold. Components that were placed with a mechanical template were measured to have an average vertical alignment within 380\,µrad, due to their glue layers. This potentially requires compensation and thus an alignment of the vertical axis in follow-up components.
We achieved this alignment by applying pressure to one edge of the component, creating a slight, intentional, uneven wedge in the glue layer, which enabled limited vertical alignment. This approach is limited in its alignment precision, and going through several iterations can reveal the trade-off between the alignments of both axes.
\subsection{High-precision beam position measurements}
\begin{figure}
            \centering
        \includegraphics[width=1\linewidth]{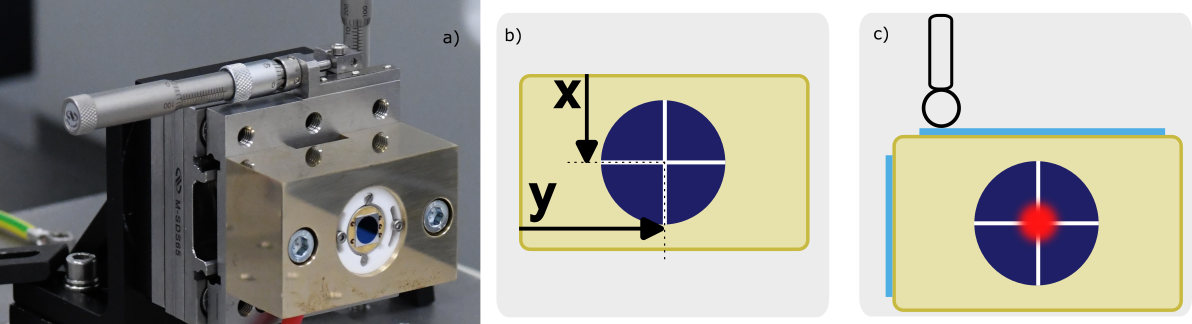}
            \caption{A photograph of a CQS is shown in a). The housing, measured by the CMM, is made of brass and mounted on two micrometer-controlled translation stages. The following two pictures illustrate the working principle of the CQS. b) A calibration procedure provides the distances between the housing to the center of the QPD. The third, z-direction, is not shown for simplicity. c) The beam is centered by using differential power sensing on a QPD, and by measuring the outside of the mount, here highlighted in blue, the position of the beam is revealed. }
    \label{fig:CQS_principle}
\end{figure}

Beam alignments with high precision require a high-precision beam position measurement tool. Inspired by the technique of a Calibrated Quadrant Photodiode Pair (CQP) \cite{FitzsimonsCQPPaper}, we developed and characterized the Calibrated Quadrant Photodiode Singleton (CQS).
As depicted in figure \ref{fig:CQS_principle} a), a large surface Quadrant Photodiode (QPD) without its protective window is attached to a cuboid brass mount. They interface via the low thermal expansion and electrically isolating material Macor. The assembly is screwed onto two translation stages, enabling fine position adjustments.
Figure\,\ref{fig:CQS_principle} further illustrates that the CQS combines the spatial position readout of a beam via Differential Power Sensing (DPS) (see e.g. \cite{Wanner2012DPS}) on a QPD with precision position measurements of the brass block by a Coordinate Measurement Machine (CMM).
A sophisticated calibration procedure (in its basic principle comparable to \cite{FitzsimonsCQPPaper}) is applied to determine the distances between the housing surfaces to the center of the installed QPD.
To measure the position of a beam, it is centered on the QPD, the housing is measured, and the calibrated values are applied.
We estimate the positional standard readout uncertainty as ±3.5\,µm and observe that the CQS stays in position over several days within 5\,µm in all axes.
Angular measurements of a beam can be performed via two spatially separated measurements as described in the next section and have a standard measurement uncertainty of ±12\,µrad.
Comparable values have been reported for CQP-based measurements, for example, ±4\,µm and ±20\,µrad \cite{FitzsimonsCQPPaper}, ±4\,µm, and ±30\,µrad \cite{DehnePhD}, and 10\,µm and ±50\,µrad \cite{LiCQPPaper}, even though not all of these were computed in the form of standard uncertainties.\\
CQSs are relatively simple to manufacture and calibrate. Working with three CQSs enabled us to develop strategies that require simultaneous beam measurements. Further advantages of the CQSs are their small footprint, which allows for flexibility and easy one-point measurements and references. This is especially useful when an optical bench has several output ports where these CQSs can be placed.
\subsection{Alignment based on CQS targets}
Besides the option of centering the CQS on a beam in order to extract its position, the process can also be reversed to align a beam to a predefined position. By using CMM measurements to place the center of the CQS at specific coordinates, the beam simply needs to be centered on the QPD to be aligned at this position. We refer to the CQS as 'target' in this process.\\
Positions for targets can, for example, be extracted from simulations of the interferometer.
We found it, however, more feasible, intuitive, and flexible to precede the alignment of the actual beam with an external, auxiliary alignment beam.
It is created by a commercial coupler and two adjustable mirrors, allowing for reproducible non-time-constrained alignment procedures until all the requirements on the beam's orientation are fulfilled.
Moreover, this strategy lends itself well to exploring different options for aligning a beam relative to the already placed components.\\
When the preferred alignment is found, two CQSs are placed at two different output ports of the optical bench with two distinct lever arms. Their two positions and thus those of the beam are measured. This 'saves' or 'records' the alignment, because the beam's vector can be entirely determined by two position measurements.\\
The advantage of the CQS, opposed to any other simple target, lies in the fact that CQS measurements are all relative to the optical bench. If any microscopic or macroscopic motion of the optical bench is expected, a simple re-measurement of the position of the optical bench can be performed. These motions can, for example, be due to temperature drifts or mechanical creep over nights and weekends, or they can be a consequence of deliberately repositioning the bench on the CMM table.
\subsection{High precision beam alignment measurements}
The orientation of a beam is measured by performing two CQS measurements at two different positions along the beam, while the alignment of a beam is done using two targets, as described in the previous section.
Most beam alignments were performed to reach a specific mode matching with other beams, where the alignment matrix is optimizing the overlap. One of two exceptions is the alignment of the beam to an absolute orientation over the optical bench, for example, focusing on a specific height.
We achieved this with typical deviations of 1\,µm $\pm$3.5\,µm and 4\,µrad $\pm$12\,µrad from the predefined target beam axes.\\
A significant challenge is aligning two counter-propagating beams to overlap with a precision of a few micrometers and microradians.
CQS measurements of both beams were used to determine the vectors of the beams. Based on those, we calculated ideal target positions and aligned the beams accordingly. By doing this in an iterative process, the final angular alignment is limited by the CQS's uncertainty of 12\,µrad.
In our construction, further constraints required a trade-off, and we settled for an angle of 27\,µrad$\,\pm\,$12\,µrad, while the beam distance at the relevant combining beamsplitters was below 5\,µm$ \,\pm\,$3.5\,µrad.
\subsection{Construction techniques for fiber couplers}
The design of the 3BL requires several fiber-to-fiber couplings, thus necessitating a sufficient mode match between the in- and out-coupled beams. This addresses the orientations of the beams as well as their beam parameters. Both are controlled during the construction of our fiber injector optical subassembly (FIOS) \cite{PenkertDiploma}, which is performed directly on the optical bench.
The FIOS design enables us to control nearly all degrees of freedom simultaneously, but has the disadvantage that they are also highly coupled and not easily reproducible.\\
We address this challenge by using the auxiliary alignment beam mentioned previously. After optimizing its alignment and recording its position and orientation via reproducible targets, the actual FIOS is placed so that its beam is centered on these two targets. It automatically fulfills all geometric requirements of the beam axis that were achieved with the alignment beam, eliminating the need for time-consuming verification measurements, which would otherwise allow temperature drifts to couple into the alignment.
Simultaneously, a beam profile camera is placed at a third output port of the bench to observe and adjust the beam's waist size and its position.
\subsection{Detailed construction example}\label{subsec:construction example}
To visualize the different strategies we developed, we present here one especially challenging and essential construction step, along with the applied techniques.
We address the challenges of aligning two FIOS, coupling light into one of them, achieving contrast, and controlling the absolute height of two beams, all simultaneously.
This concerns the components labeled in figure\,\ref{fig:3BLoverview} by FIOS4, bs12, FIOS3 and bs11.
\begin{figure}[ht]
    \centering
    \includegraphics[width=0.95\linewidth]{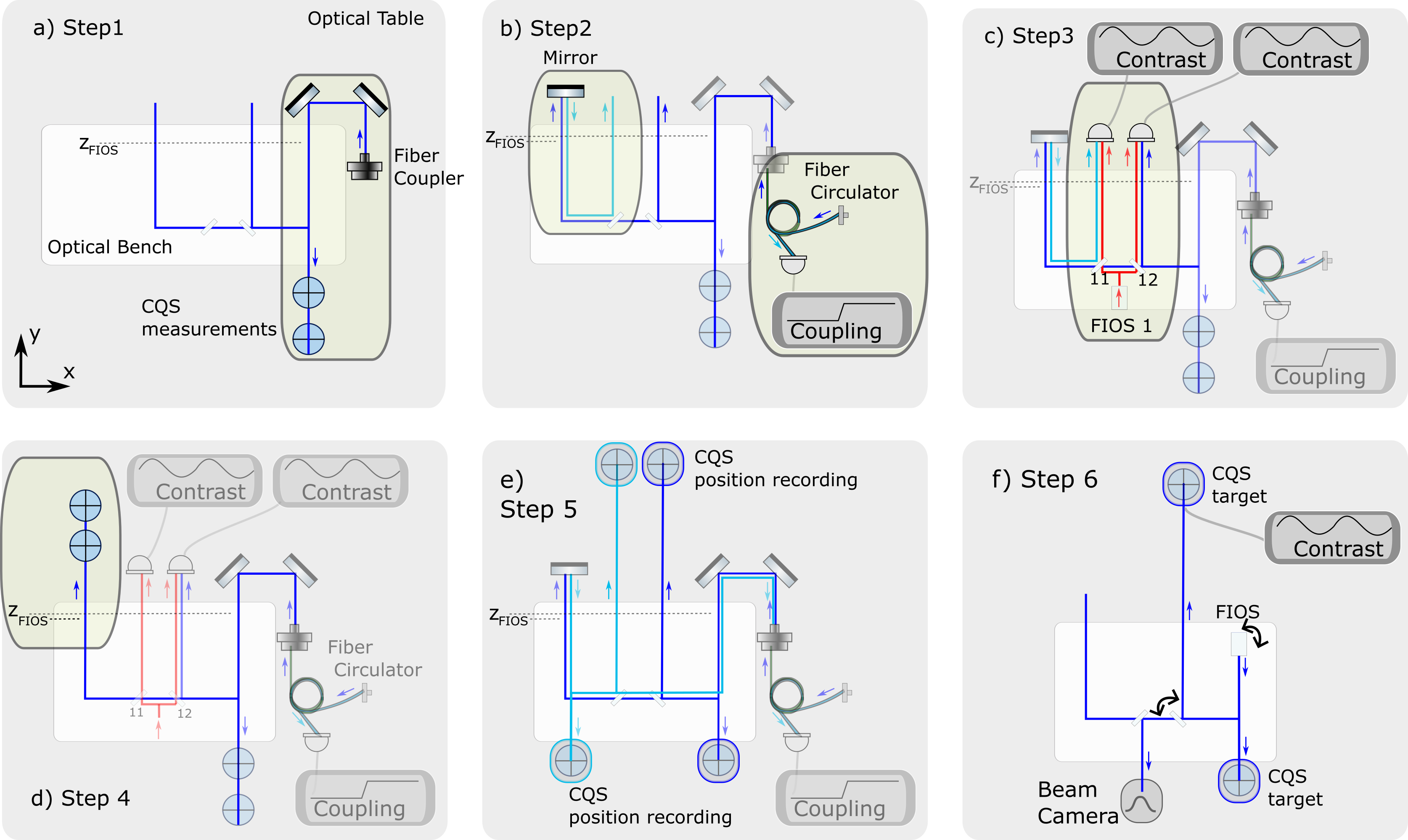}
    \caption{Step-by-step illustration of the construction example. Note that the optical bench layout has been significantly simplified compared to the actual design in figure\,\ref{fig:3BLoverview}. a) The first alignment beam is created by commercial components and aligned according to CQS measurements. b) A second alignment beam is created by back-coupling into the fiber coupler. c) The two bs are placed and aligned in order to optimize the contrast. d) The mirror is removed, and by CQS measurements, the height at the FIOS position is estimated. e) A pair of CQS measurements is performed on each alignment beam. f) One of the FIOS is placed and aligned according to the CQS measurements. This figure is based on a similar figure in \cite{BischofPhD}.}
    \label{fig:secenario4Pic}
\end{figure}
\subsubsection*{Step 1}
An alignment beam (dark blue) is created by a commercial coupler and two mirrors as shown in figure~\ref{fig:secenario4Pic}~a). Two CQS measurements along one beam path enable the calculation of the orientation and absolute height of that beam with respect to the optical bench. The focus is on a horizontal beam alignment and the beam height at the FIOS position ($z_{\text{FIOS}}$). This is crucial because the height of the FIOS output at 15\,mm is not adjustable.
\subsubsection*{Step 2}
Figure~\ref{fig:secenario4Pic}\,b) illustrates the process of handling the FIOS-to-FIOS coupling.
A commercial mirror under normal incidence is placed behind the bench at the x-coordinates of the second FIOS. 
It is aligned to maximize the coupling of the light back into the fiber coupler. Hereby, the reflected beam mimics the second FIOS's output, creating an additional alignment beam (light blue). 
\subsubsection*{Step 3}
Both counter-propagating, mimicked FIOS beams are now present in their respective interferometer, where they need to interfere with the already placed FIOS1 beam - illustrated in red in figure \ref{fig:secenario4Pic}\,c). The two recombining beamsplitters are temporarily placed on the bench, and the interference is optimized by optimizing the contrast observed with a photoreceiver and an oscilloscope.\\
In this specific example, the vertical alignment can be fine-tuned by adjusting the commercial mirrors of the two alignment beams; however, a trade-off between the two interference contrasts, the height at the FIOS4 position, and the incoupling must be made.
\subsubsection*{Step 4}
Once the aforementioned trade-off is found, the zero-degree mirror is removed, and the beam in that path is measured with two consecutive CQS measurements, as illustrated in figure\,\ref{fig:secenario4Pic}\,d). These measurements are used to calculate the height ($z_{\text{FIOS}}$) at the FIOS3 position. If it exceeds the FIOS height requirement of 15\,mm, the whole process has to be restarted at Step 1.
\subsubsection*{Step 5}
When all the requirements for fiber coupling, contrast, and FIOS heights are fulfilled, the next step is to record the alignment with CQS measurements, as illustrated in figure\,\ref{fig:secenario4Pic}\,e).
For each mimicked FIOS beam, two output ports of the bench are chosen, and the beam's positions are measured. The measurement of the position of the CQSs (opposed to just placing them), is required because the described Steps 1-4 as well as the placement of the FIOS require more than a day. By measuring the position of the CQSs at the end and beginning of each day, we exclude any misalignment due to temperature and other external influences.
\subsubsection*{Step 6}
As illustrated in figure\,\ref{fig:secenario4Pic}\,f), the first of the two FIOS is placed and glued on the optical bench in this step. It is centered on the two targets identified in Step 5, and the beam's waist and position are simultaneously aligned by observation on a beam camera in a third output port.
In this construction example, the corresponding recombining beamsplitter is simultaneously aligned and glued for optimized contrast. The glue of both components is cured once the centering on the targets and the optimization of the contrast are recreated from Step 3 simultaneously.
\subsubsection*{Step 7}
To conclude the alignment, the second FIOS and its beamsplitter are placed according to the procedure in Step 6.\\
\\
This approach was successfully applied for the construction of these four components on both benches. On the first bench, a sufficient trade-off between all the parameters was not possible, limited by the height ($z_{\text{FIOS}}$) at the FIOS3 position. As a solution, we were able to shorten the holder of that FIOS according to the measured amount.
The resulting coupling, contrast, and beam parameter values are discussed in table\,\ref{tab:construction_characterization}. 
\section{Characterization of the finalized construction}
\begin{figure}[ht]
    \centering
    \includegraphics[height=0.2\textheight]{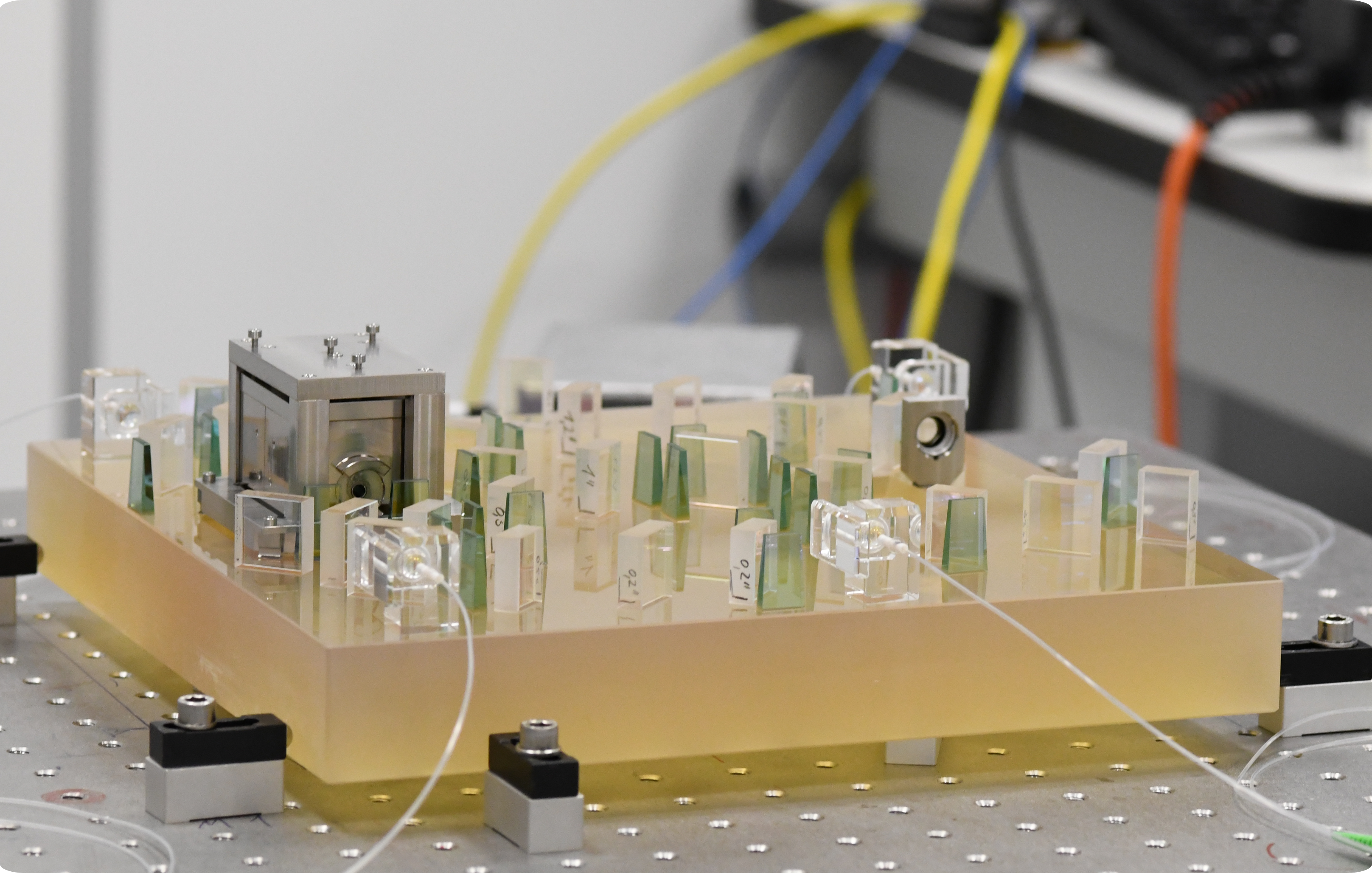}
    \includegraphics[height=0.2\textheight]{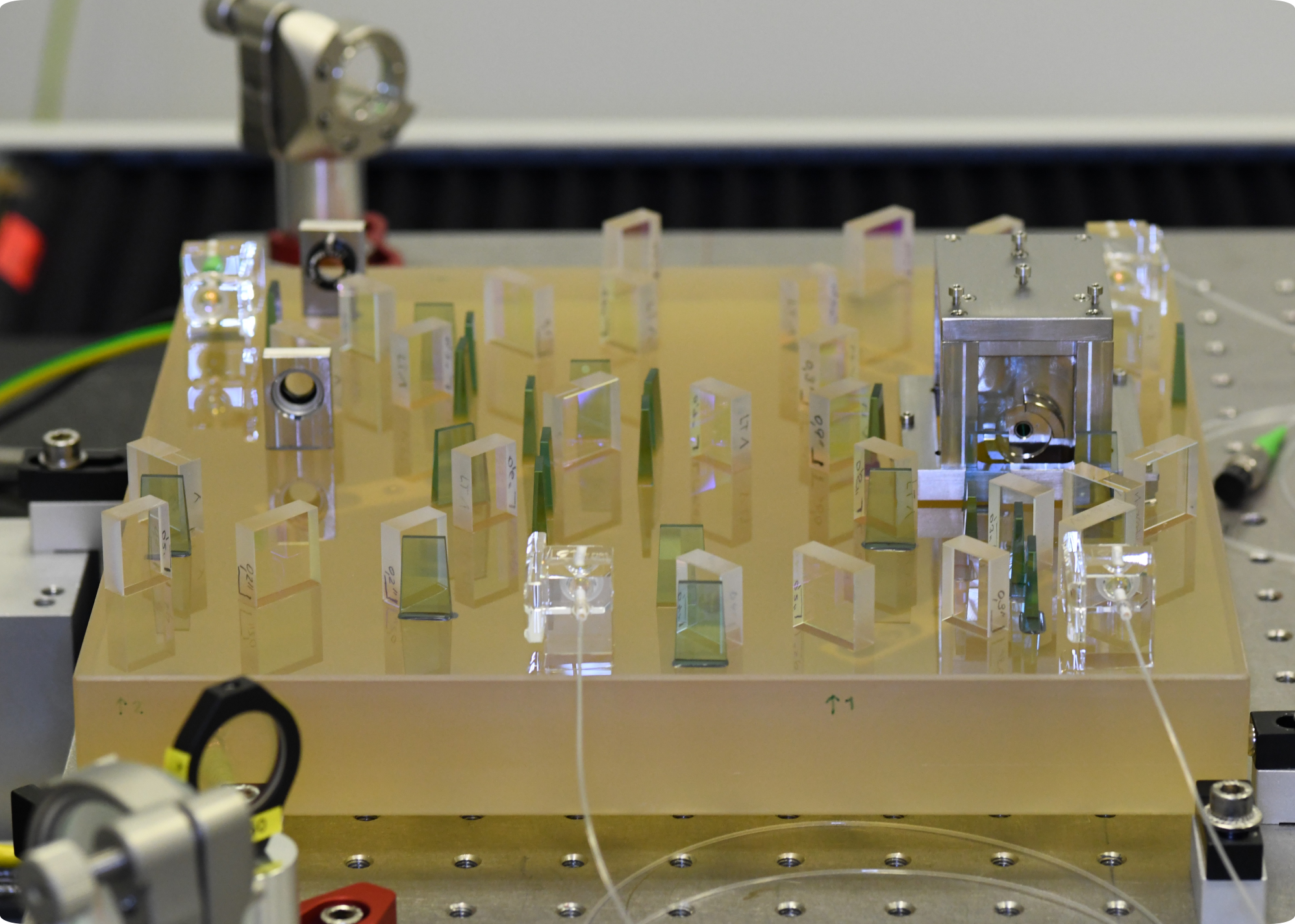}
    \caption{Pictures of the finalized optical benches that form the 3BL experiment. All components were placed according to the design, with only minor deviations in their absolute positions. The pictures were first printed in \cite{BischofPhD}}
    \label{fig:3BLBenches}
\end{figure}

The two finalized, fully functional benches are shown in figure\,\ref{fig:3BLBenches}.
\subsection{Contrast, coupling and beam parameters}
During and after construction, the most relevant parameters of the interferometers were characterized and summarized in table\,\ref{tab:construction_characterization}.
The contrast is defined with the heterodyne efficiency $\eta$ and the powers of the two interfering beams within one interferometer $P_1$ and $P_2$ by $$c=2\frac{\sqrt{P_1P_2}}{P_1+P_2}\sqrt{\eta}=\frac{V_\text{min}-V_\text{max}}{V_\text{min}+V_\text{max}},$$
and measured in the laboratory with a photodiode equipped with a transimpedance amplifier by the minimal and maximal voltage levels $V_\text{min}$, $V_\text{max}$, on an oscilloscope. The given contrast values were measured for approximately equalized beam powers and are thus representative of the heterodyne efficiencies.\\
The coupling into the Backlinks is defined as the ratio of incident to outcoupled power.
\begin{table}[ht]
    \centering
    \begin{tabular}{lc||lcc||lc}
        \textbf{Interferometer} & contrast & \textbf{FIOS} & $\omega_0$ & z$(\omega_0)$ & \textbf{Backlink} & coupling\\
         & [\%] &  & [µm] & [mm] &  & [\%]\\
        \hline
        DFBL Bench 1 & 71 & FIOS1 Bench 1 & 450 & 501 & DFBL Bench 1 & 52\\
        DFBL Bench 2 & 58 & FIOS1 Bench 2 & 371 & 450 & DFBL Bench 2 & 50\\
         \hline
         FSFBL-meas Bench 1 & 52 & FIOS2 Bench 1 & 578 & 207 & FSFBL Bench 1 & 68 \\
         FSFBL-meas Bench 2 & 90 & FIOS2 Bench 2 & $\approx$500 & $\approx$ 0 & FSFBL Bench 2 & 26\\
                  \hline
         FSFBL-ref Bench 1 & 74 & FIOS3 Bench 1 & 579 & 647 &  & \\
         FSFBL-ref Bench 2 & 83 & FIOS3 Bench 2 & 350 & 550 &  & \\
                  \hline
         FBBL Bench 1 & $\approx$65 & FIOS4 Bench 1 & 529 & 201 &  & \\
         FBBL Bench 2 & $\approx$65 & FIOS4 Bench 2 & 427 & 512 & &\\
    \end{tabular}
    \caption{Characterization parameters of optical benches 1 and 2. Note that the FBBL has no fiber-coupling value, and its measured contrast during construction is not fully representative of the final interference using the steering mirrors in the experimental installation.}
    \label{tab:construction_characterization}
\end{table}
The coupling values achieved during the construction step explained in \ref{subsec:construction example} were at more than 60\,\% on both benches. However, on the second bench, a degrading glue-layer reduced this to less than 30\,\% after the finalization of the construction step.
This low coupling of 26\% into one of the FSFBL reduces the amount of light that is transmitted to the other bench. It results in a reduced power ratio between the two interfered beams, which is at this level still uncritical for the coupling of additive noises like RIN \cite{IsleifPhD}.
\subsection{Component Tilts}
In components that were not monitored or controlled in their vertical axis, we observe tilts in the range between 100 to 700\,µrad.
Intentionally induced tilts are on the same order of magnitude, except one, where a significant tilt of 3\,mrad was required for a satisfactory alignment of the beam.\\
We estimate here the effect of the thermal expansion of the glue layer, which is related to the 3\,mrad tilt, on a pathlength change of the reflected beam.
The bottom side of the components has an area of 7\,x15\,mm$^2$ and a glue amount of 0.01\,µl/mm$^2$ was used.
The dominant effect on the pathlength is here described by the piston effect \cite{SchusterPhD}. By combining this with additional geometry and the CTE of the glue (55$\cdot10^{-6}$/K \cite{OptocastDatasheet}) we derive a conservative upper limit for the coupling from temperature changes to pathlength changes of:
$$
\frac{ds}{dT}=10\frac{\text{nm}}{\text{K}}.
$$
With the thermal stability in our vacuum chamber that reaches a noise floor of $1\cdot10^{-5}\frac{\text{K}}{\sqrt{\text{Hz}}}$, this translates to a noise floor of $0.1\frac{\text{pm}}{\sqrt{\text{Hz}}}$.
Accordingly, the significant tilt of 3\,mrad is not expected to limit the optical performance.
\section{Initial performance of the Three-Backlink Experiment}
In this section, we present the embedding of the 3BL benches in the laboratory infrastructure.
This is followed by the results of the non-reciprocity measurements performed after an initial commissioning phase. More information about the infrastructure preparation and these inital measurements can be found in \cite{JestrabekMaster} and \cite{BischofPhD}.
\subsection{Laboratory infrastructure}
To probe the Backlinks for their noise performance in the $\frac{\text{pm}}{\sqrt{\text{Hz}}}$ range, their surrounding vacuum chamber and electronics must provide a sufficiently low-noise environment. Likewise, the read-out and control system have to enable verification and monitoring of this low noise level. Many aspects of this infrastructure have already been described in \cite{Isleif2018}.
\subsubsection{Inside the vacuum chamber}
\begin{figure}[ht]
    \centering
    \includegraphics[width=0.95\linewidth]{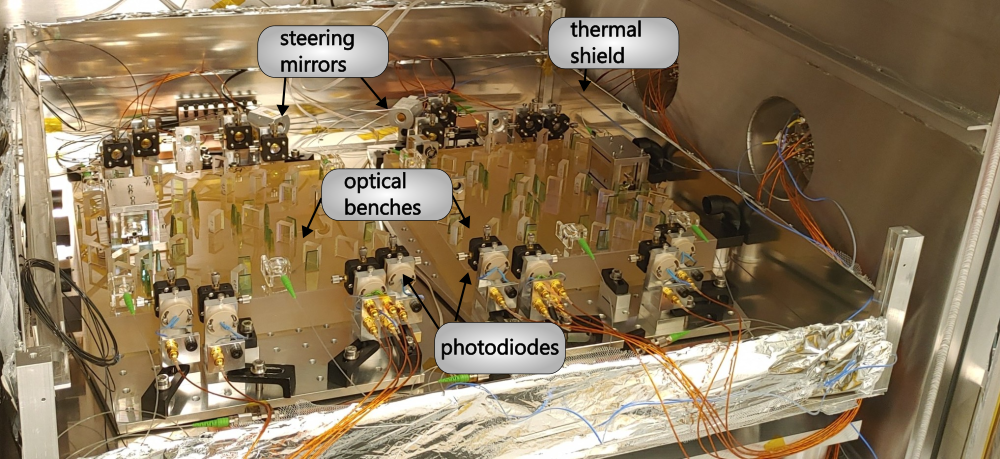}
    \caption{The 3BL optical benches are installed inside the vacuum chamber. Photodiodes are installed at the front and the back, and the FBBL-steering mirrors are aligned at the back. Parts of the thermal shield have been removed for visibility.}
    \label{fig:3BLinChamber}
\end{figure}

Figure\,\ref{fig:3BLinChamber} shows the two optical benches placed inside a vacuum chamber.
Each bench is placed on an aluminum breadboard that houses the steering mirrors for the FBBL, S-334 by Physics Instruments.
Next to the mirrors are the photodiodes placed at both output ports of each interferometer, and for the power pick-off for laser power stabilization. Each diode holder is further equipped with a lens to enable focusing on the small diode and a polarizer to filter unwanted polarization effects.
Only the diodes (single-element InGaAs 1\,mm diameter and quadrant InGaAs GAP1000Q) are placed inside the vacuum chamber, while their transimpedance amplifiers (TIA) are located on the outside. This design was chosen due to the high number of required photoreceivers and their resulting thermal load.
The system is enclosed in a thermal shield made of aluminum wrapped in insulation foil.
Through the bottom layer of the thermal shield, as shown in figure\,\ref{fig:experimentalInfrastructure}, the aluminum breadboards are mounted on a rotation stage each. These vacuum-compatible stepper motors by Newport enable a rotation between the benches.
Vibration dampers isolate the system against outside disturbances and thermal sensors on the benches and fibers monitor the thermal environment, which reaches a noise floor of $ 1\cdot 10^{-5}\frac{\text{K}}{\sqrt{\text{Hz}}}\cdot\sqrt{1+\frac{(2.0\text{mHz})^4}{\text{f}}}$.
A combination of pre-and turbo pumps enables us to reach a pressure level around $1\cdot10^{-5}$mbar. 
\subsubsection{Outside the vacuum chamber}
\begin{figure}[ht]
    \centering
    \includegraphics[width=0.95\linewidth]{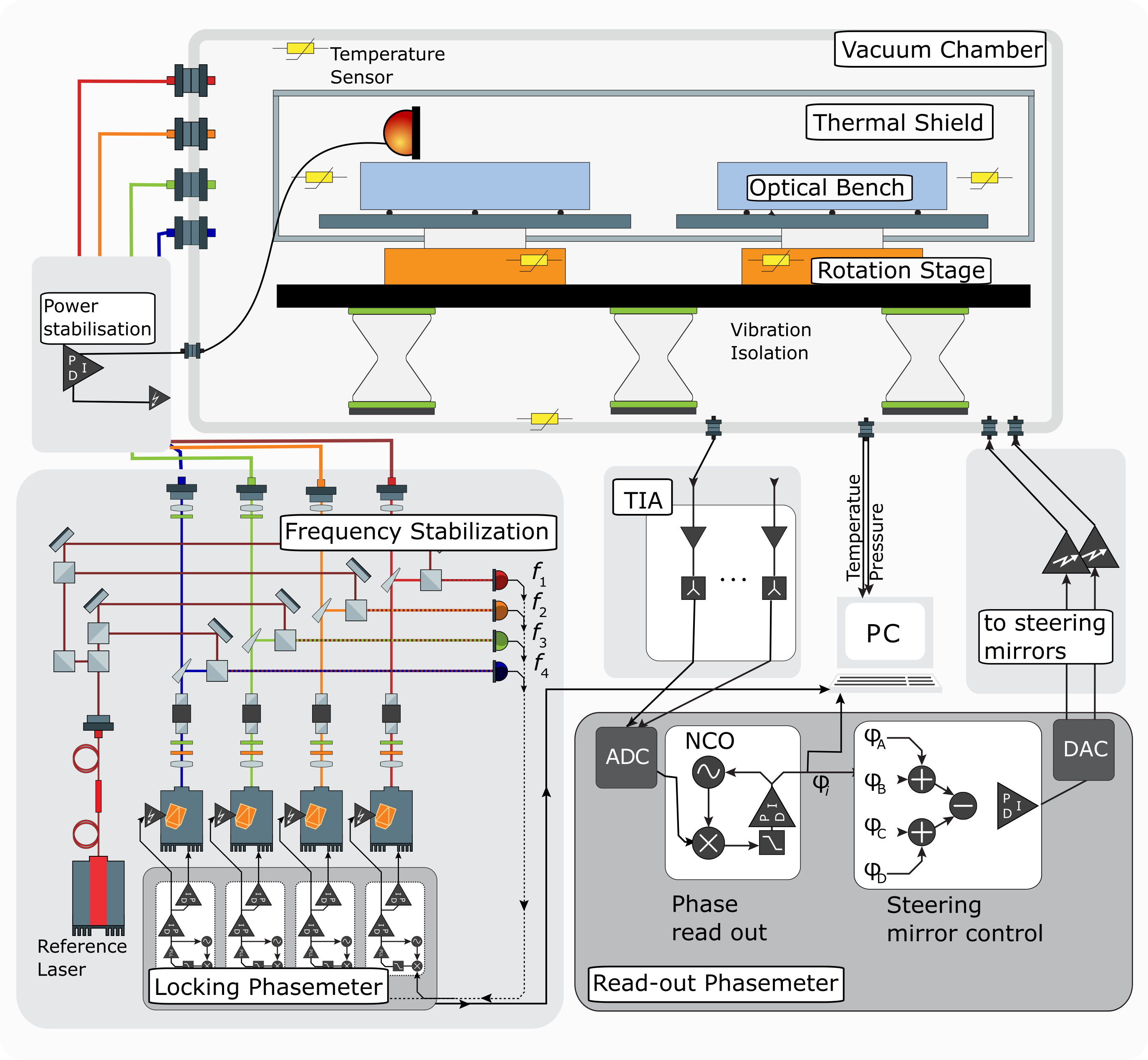}
    \caption{Experimental overview of the 3BL infrastructure. On the lower left, the four lasers are stabilized in frequency to a reference laser, creating the desired beatnote frequencies in the experiment. In the top part, the vacuum chamber housing the optical benches is shown, and in the lower right, the read-out and steering mirror control system is depicted. Figure based on a figure in \cite{IsleifPhD}.}
    \label{fig:experimentalInfrastructure}
\end{figure}

Figure\,\ref{fig:experimentalInfrastructure} shows a schematic overview of the surrounding support equipment.
The frequencies of the four lasers are offset-locked to a stable reference laser by means of a custom-built Phasemeter. The latter consists of an ADPLL (all-digital phase-locked loop) \cite{GerberdingPhD}, serving as a frequency sensor, and digital controllers that act on the laser crystal temperature and cavity length \cite{IsleifPhD}. The four offset lock values are within the MHz range and can be adapted conveniently to achieve kHz beatnotes between the individual lasers.
The reference laser is a Prometheus laser by Coherent, which enables internal stabilization to an iodine cell and a laser frequency noise below $300\frac{\text{Hz}}{\sqrt{\text{Hz}}}$ with the relevant noise shape \cite{LieserPhD}.
After polarization cleaning in front of the feedthrough fibers, the four laser beams are power-stabilized with a control loop that acts on variable optical attenuators using the power pick-off photodetectors inside the vacuum chamber.
The readout of the setup is illustrated on the right side in figure\,\ref{fig:experimentalInfrastructure}, where the signals from the measurement photodetectors are sent to the TIAs and are then fed into the in-house built readout phasemeter.
This readout phasemeter is based on an IQ-demodulation \cite{GerberdingPhD}. It is modified with a low-bandwidth feedback mechanism between the Q output and the local oscillator \cite{RischkopfMaster}. This ensures tracking of the beatnote phase despite transient spikes, e.g., caused by the discrete movement of the stepper motor. Different readout filters are used to generate scientific data and sensor data for steering mirror control in parallel.
\subsection{Results}
\begin{figure}[ht]
    \centering
    \includegraphics[width=0.99\linewidth]{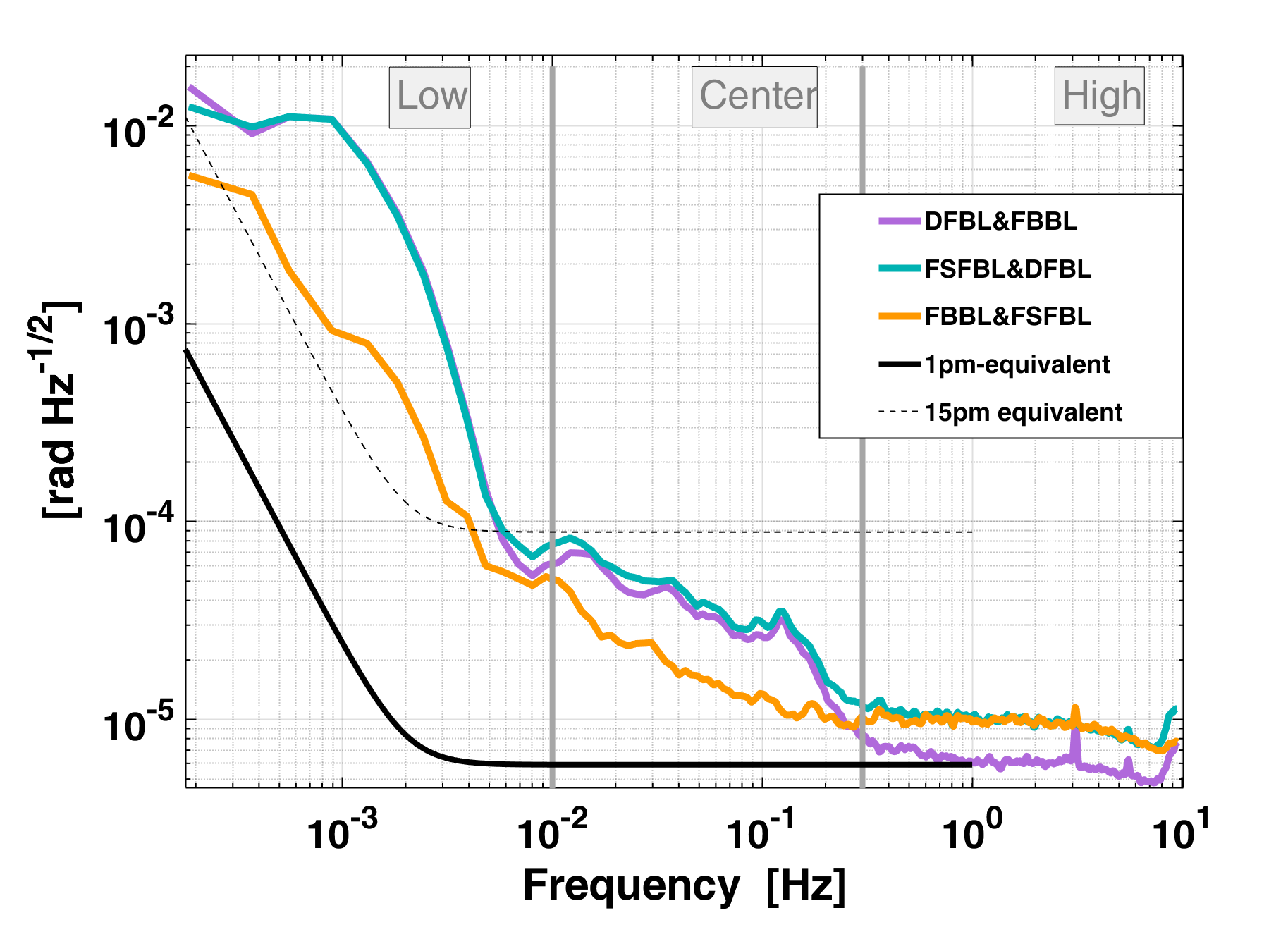}
    \caption{Non-reciprocal noise level of the three Backlink pairs. The 1\,pm-equivalent noise level is given as reference. A 15\,pm-equivalent noise illustrates the achieved upper limit. The figure is a re-print with several cosmetic changes from \cite{BischofPhD}.}
    \label{fig:3BLresultFirstLast}
\end{figure}

Figure\,\ref{fig:3BLresultFirstLast} shows the results of the first 3BL measurements. Three curves, illustrating the three pairwise combinations presented in equation~\ref{eq:All3BLsEquation}, are compared to a 1\,pm-equivalent-noise floor, multiplied with the noise shape function: $\text{req}(\text{f})=1\text{pm}\cdot\frac{2\pi}{\lambda}\cdot\sqrt{1+\frac{(2.0\text{mHz})^4}{\text{f}}}$.
The shown frequency range from 10\,Hz to 0.1\,mHz is divided into three ranges, which allow for a more distinguished description in the following sections where we describe improvements and residual limitations. In general, operating in vacuum conditions is essential to achieve the presented noise levels in the center and low-frequency range.\\
In the low and center frequency ranges, the curves of DFBL\&FBBL and FSFBL\&DFBL overlap. From this observation, we conclude that noise related to the DFBL implementation limits both combinations. In the high-frequency range, we can use the same argument to see that noise related to the FSFBL is limiting.
We discuss in the following how we can relate certain limitations to the laboratory infrastructure.
\subsubsection{High-frequency range - above 0.3~Hz}
The noise levels of the non-reciprocity measurements in the high frequency range show a limitation by a white noise level at $1\cdot10^{-5}\frac{\text{rad}}{\sqrt{\text{Hz}}}$ and for one combination down to the requirement of $6.6\cdot10^{-6}\frac{\text{rad}}{\sqrt{\text{Hz}}}$.
The limiting noise is a beatnote-frequency dependent noise level, presumably caused by the phase read-out system. 
The exact cause could not be identified, but the most likely candidates are a parasitic signal at low kHz frequencies inside the phase read-out system or the high phase dynamics of the optical signals.
The latter would couple via the modified IQ-demodulation of the phasemeter, which includes a Cascaded Integrator-Comb (CIC) filter. The beatnote frequencies are chosen to exploit notches in its transfer function for high suppression of undesired harmonics. However, excessive dynamics of the beatnotes might push these undesired signals into a domain around the notches where suppression becomes insufficient. In that case, the transfer function envelope becomes more relevant, which has a higher suppression at higher frequencies.\\
We found that frequencies of a few 100\,kHz instead of frequencies in the order of 10\,kHz resulted in a lower non-reciprocity noise floor.
The Backlink pair reaching the 1\,pm-equivalent requirement (DFBL\&FBBL) operates with two lasers and thus one beatnote frequency that can be chosen freely.
The FSFBL, and thus its combinations, requires four additional frequencies. Due to their entanglement, not all frequencies can be chosen in the optimal range, and the remaining noise floor is limited by this beatnote-frequency dependent phase read-out noise.
 \subsubsection{Center-frequency range - 0.01~Hz to 0.3~Hz}
In this frequency range, we observed an improvement in the FSFBL by implementing the laser power stabilization, which is activated for the measurement shown in figure\,\ref{fig:3BLresultFirstLast}. More detailed investigations of the coupling mechanism were performed later and are presented in \cite{Ho-Zhang2025}.\\
The combination of FBBL\&FSFBL shown in figure\,\ref{fig:3BLresultFirstLast} has the lowest noise level of the three, with a rise to lower frequencies. Which one of the two Backlink implementations in this combination is limiting cannot be identified from this plot.
However, we conclude that the other two combinations, shown in the overlapping green and violet lines, are limited by the DFBL. Especially prominent is a shoulder around 0.25\,Hz that will be discussed in the following.\\
We assume that the limiting noise is caused by a coupling via fiber backscatter, based on our knowledge from \cite{FleddermannPhD} and \cite{RohrPhD} about the significant backscatter influence in this Backlink.
Its coupling to the phase measurement can occur via two effects: one is the thermal expansion of the fiber driving the phase dynamics, and the other is laser frequency noise coupling via the additional pathlength difference of the scatter.
We further conclude that the latter effect is dominant in this frequency range, based on several observations presented below.\\
While LFN couples to a phase measurement via $\varphi_{\text{error}}=\frac{2\pi}{c}\cdot\Delta\text{s} \cdot\text{LFN}$, this cancels for perfectly symmetrical benches, as we had assumed in equation\,\ref{eq:non-recDFBL}.
Accordingly, if we do not use the sum as in equation\,\ref{eq:non-recDFBL}, but the difference, we get a phase that includes twice the LFN coupling in both directions. By combining it with the known pathlength differences in the interferometer ($\Delta\text{s-DFBL}=2.3$m), we get a measurement of the LFN of the involved lasers.
We compared previous NPRO LFN measurements to the one we just described from the phase measurements and found that they match well in shape and level. 
The reason that the LFN matches the one of an NPRO, instead of the above mentioned high frequency stability, lays in technical issues: The reference laser could not be stabilized on its iodine reference and instead exhibited frequency noise of a free running NPRO of up to $1\cdot10^6\frac{\text{Hz}}{\sqrt{\text{Hz}}}$ at 0.1Hz. Therefore, the LFN was significantly higher than what the laboratory infrastructure was designed for.\\
The shape of this calculated LFN furthermore matches the 0.25\,Hz shoulder in the non-reciprocity and thus implies a linear coupling mechanism.
While this is not a definite validation, we find an additional argument for backscatter-related coupling by applying balanced detection. This technique, which combines measurements at both output ports of each DFBL interferometer, subtracts the contributions of backscattered light from the measurement (as discussed, for example, in \cite{FleddermannPhD}). This post-correction method indeed reduced the shoulder at 0.25\,Hz by about a factor of 2, making it very likely that the reduced noise was caused by backscatter. Note that the resulting, improved noise curve is not shown in figure\,\ref{fig:3BLresultFirstLast} because we want to focus on results obtained without post-correction methods.\\
Still, we used the noise level with balanced detection to further investigate the underlying noise floor. Here, we assume another coupling mechanism of LFN. Since the aforementioned, previously assumed perfect symmetry of the optical benches is not realistic, residual asymmetries will also lead to an LFN coupling.\\
As mentioned earlier, at the time of these measurements, the LFN could not be improved due to technical reasons, and accordingly, the presented assumptions could not be experimentally verified. Accordingly, we provide initial evidence of verifying the noise coupling via LFN, while a more detailed analysis and the expected improvement in non-reciprocities with improved LFN are discussed in \cite{Ho-Zhang2025}.\\
It should further be noted that the here-discussed LFN noise coupling is not a limitation for the DFBL design, but rather an example of how the 3BL can be used to study different noise contributions and coupling mechanisms in the future.
\subsubsection{Low-frequency range - below 0.01~Hz}
In the DFBL, the second component that drives the backscatter phase dynamics is the thermal expansion of the fiber. Temperature effects typically become visible in the low-frequency range, which thus likely couple into our phase measurements here.\\
Based on these measurements by the 3BL, non-linear coupling effects of temperature were studied further as part of ongoing work. These observations have been used to formulate a temperature drift requirement for LISA, verifying the 3BL as a valuable testbed for the Backlink.
The limiting noise source of the other combination could not be identified at this point.
 \section{Conclusion}
In this paper, we present the 3BL as a testbed for investigating Backlink connections. After the design choices and preparations have been discussed in \cite{Isleif2018}, we provide a condensed design overview here that focuses on the extraction of the non-reciprocity performance. 
We present the techniques and, by one example, the sophisticated alignment strategies that were required to construct the ultra-stable optical benches. These further include the CQS as a key tool for our beam measurements and alignments with µm-precision.\\
The finalized benches serve as a testbed for comparing three different Backlinks implementations: a free beam and two fiber connections. One of the two fiber designs closely resembles the LISA baseline and thus enables studying noise sources and coupling mechanisms that are inherent to a direct fiber design.
Furthermore, all three Backlink implementations are relevant to other future missions, such as Taiji and TianQin, which have not yet established a mission baseline.\\
In the last chapter of this paper, we present the measurements performed after initial commissioning. A non-reciprocal noise of 15\,$\frac{\text{pm}}{\sqrt{\text{Hz}}}\cdot\sqrt{1+\frac{(2.0\text{mHz})^4}{f}}$ was reached in the high and center frequencies. This noise level was attributed to the test environment and can be regarded as an upper limit for the stability of the optical benches as well as the performance of the three Backlink implementations.\\
The next step is to minimize the influence of the test environment to achieve a $1\frac{\text{pm}}{\sqrt{\text{Hz}}}$ performance level. This work is presented in \cite{Ho-Zhang2025}, which finalizes the preparation of the 3BL as Backlink testbed. The following work will utilize this testbed to apply rotation between the benches, perform comparative analyses of the three Backlinks, and study noise coupling mechanisms in detail.
\section*{Author contributions}
Conceptualization: Gerhard Heinzel, Katharina-Sophie Isleif, Oliver Gerberding, Karsten Danzmann\\
Formal analysis Laboratory: Melanie Ast, Lea Bischof\\
Funding acquisition: Jens Reiche \\
Investigation Construction: Lea Bischof, Stefan Ast, Nicole Knust, Daniel Penkert, Katharina-Sophie Isleif, Oliver Gerberding\\
Investigation Laboratory: Lea Bischof, Melanie Ast, Daniel Jestrabek, Jiang Ji Ho-Zhang\\
Methodology: Lea Bischof, Daniel Penkert, Oliver Gerberding, Katharina-Sophie Isleif, Nicole Knust, Stefan Ast\\
Methodology CQS: Daniel Penkert, Lea Bischof\\
Project administration:  Gerhard Heinzel, Jens Reiche\\
Software Laboratory: Thomas S. Schwarze, Oliver Gerberding\\
Software Construction: Daniel Penkert, Lea Bischof\\
Supervision:  Stefan Ast, Oliver Gerberding\\
Visualization: Lea Bischof\\
Writing–original draft: Lea Bischof\\
Writing–review \& editing:  Lea Bischof, Melanie Ast, Jiang Ji Ho-Zhang\\
\section*{Acknowledgments}
We gratefully acknowledge support by the European Space Agency (ESA) within the project "Phase Reference Distribution System" (8586/16/NL/BW) and the Deutsches Zentrum für Luft- und Raumfahrt (DLR) with funding from the Bundesministerium für Wirtschaft und Technologie (Project Ref. Number FKZ50OQ1801).\\
We gratefully acknowledge the laboratory support of our colleagues Max Rohr, Michael Born, Christoph Bode and Juliane von Wrangel.
We gratefully acknowledge fruitful discussions with Alasdair Taylor and David Robertson from the University of Glasgow.
Further thanks goes to the great support structure at AEI, including especially the mechanical and electronic workshop.\\
We also acknowledge previous studies of the LISA Backlink at the AEI, including work by Roland Fledderman, Christian Diekmann, Frank Steier and Jan-Simon Henning.
\section*{Data}
Data generated or analyzed during this study are available from the corresponding author upon reasonable request.
\printbibliography
\end{document}